%% file: main.tex
\documentclass[preprint,12pt,numafflabel]{elsarticle}
\pdfoutput=1

\usepackage{amsmath,amssymb,amsfonts}
\usepackage{algorithmic}
\usepackage{graphicx}
\usepackage{textcomp}
\usepackage[table]{xcolor}
\usepackage{booktabs}
\usepackage{multirow}
\usepackage{subcaption}
\usepackage[most]{tcolorbox}
\usepackage{verbatim}
\usepackage{url}
\usepackage{hyperref}
\usepackage{enumitem}
\usepackage{threeparttable}

\definecolor{lightgreen}{rgb}{0.8,1,0.8}
\definecolor{summarybox}{rgb}{0.9,0.95,1}
\definecolor{summaryheader}{RGB}{0,51,153}

\newcommand{\RqOne}{\textbf{RQ1:} \emph{How do repositories with ACFs differ from repositories without ACFs?}}
\newcommand{\RqOneOne}{\textbf{RQ1.1:} \emph{How do the two repository groups differ in the types of software?}}
\newcommand{\RqOneTwo}{\textbf{RQ1.2:} \emph{How do the two repository groups differ in README structural characteristics?}}
\newcommand{\RqTwo}{\textbf{RQ2:} \emph{How do the issues in repositories with ACFs differ from the issues in repositories without ACFs?}}

\input{comments}
\providecommand{\tsuchida}[1]{#1}
\renewcommand{\fixme}[1]{#1}

\journal{Journal of Systems and Software}

\begin{document}

\begin{frontmatter}

\title{\tsuchida{Repository and Issue Characteristics of Early GenAI Adopters on GitHub}}

\author[1]{Rikuto Tsuchida}
\ead{tsuchida.rikuto.tq5@naist.ac.jp}

\author[1]{Youmei Fan\corref{cor1}}
\ead{fan.youmei.fs2@is.naist.jp}

\author[2]{Kazumasa Shimari}
\ead{shimari@wakayama-u.ac.jp}

\author[3]{Raula Gaikovina Kula}
\ead{raula-k@ist.osaka-u.ac.jp}

\author[4]{Gema Rodr\'iguez-P\'erez}
\ead{gema.rodriguezperez@ubc.ca}

\author[1]{Kenichi Matsumoto}
\ead{matumoto@is.naist.jp}

\cortext[cor1]{Corresponding author}

\affiliation[1]{organization={Nara Institute of Science and Technology},
            addressline={8916-5 Takayama-cho},
            city={Ikoma},
            postcode={630-0192},
            state={Nara},
            country={Japan}}

\affiliation[2]{organization={Wakayama University},
            addressline={930 Sakaedani},
            city={Wakayama},
            postcode={640-8510},
            state={Wakayama},
            country={Japan}}

\affiliation[3]{organization={The University of Osaka},
            addressline={1-5 Yamadaoka},
            city={Suita},
            postcode={565-0871},
            state={Osaka},
            country={Japan}}

\affiliation[4]{organization={University of British Columbia, Okanagan Campus},
            addressline={3333 University Way},
            city={Kelowna},
            postcode={V1V 1V7},
            state={BC},
            country={Canada}}

\begin{abstract}
Generative artificial intelligence (GenAI) \tsuchida{has introduced new forms of AI assistance into software development, but less is known about the repositories and issues of early adopters of GenAI tools}. We study \tsuchida{the repository and issue characteristics of early adopters who state interest in GenAI on their public profiles} on GitHub by analyzing \tsuchida{531 early adopters, 100 of their repositories with AI-tool configuration files (ACFs), 100 matched control repositories without ACFs, and 370 issues from the two repository groups}.
\tsuchida{Repositories with ACFs show modest distributional differences in selected README structural characteristics. They} contain longer README files with more code blocks \tsuchida{and include a license more often}, while \tsuchida{repositories without ACFs} contain more external URLs. \tsuchida{The category distributions of the two repository groups show weak evidence of a difference. Once GenAI relatedness is held fixed, we detect no difference in issue Type or Outcome between the two repository groups. Our data, collected from May to August 2025, record how early adopters developed software in Software Engineering 2.0, the stage in which developers use both AI assistants and autonomous AI agents. When these agents become the norm, later studies can compare their data with ours.}
\end{abstract}


\end{frontmatter}

\section{Introduction}
\label{sec:introduction}

Generative artificial intelligence (GenAI) has introduced new forms of AI assistance into software development. In current practice, developers selectively incorporate AI support while continuing to drive, review, and maintain the work.
These tools fall into two broad categories.
AI assistants such as GitHub Copilot\footnote{\url{https://github.com/copilot}} and ChatGPT\footnote{\url{https://chatgpt.com}} support developers who still write and review code themselves, accepting or rejecting suggestions as they go.
Autonomous AI agents such as Devin\footnote{\url{https://devin.ai/}} and Claude Code\footnote{\url{https://www.claude.com/product/claude-code}} instead aim to carry out development tasks end-to-end with minimal human intervention. This shift has been described as agentic software engineering~\cite{hassan2026agenticse,roychoudhury2025agenticai}.

Since the public release of ChatGPT in November 2022, GenAI tools have been widely adopted\tsuchida{~\cite{github2024survey}}.
Research on the benefits of GenAI is still at an early stage.
Prior studies have examined model output and performance~\cite{li2025rise,watanabe2026agentic}, code-completion quality~\cite{elhaji2024copilot}, and developer practices through surveys and organizational case studies~\cite{sergeyuk2025using,li2024aitool,liang2024large}.
Most software development today still falls between the assistant mode and the agent mode of AI support. We refer to this stage as \textit{Software Engineering 2.0}~\cite{hassan2026se3}, the transitional phase before fully autonomous agents (Software Engineering 3.0~\cite{hassan2026se3}) become the norm.
Less is known, however, about \tsuchida{the repositories and issues of early adopters of GenAI tools}. We therefore ask:
    \textit{\tsuchida{What are the repository and issue characteristics of early adopters who state interest in GenAI on their public profiles?}}

To answer this question, we focus on \tsuchida{GitHub users who state interest in GenAI on their public profiles}.
\tsuchida{These users state their interest in GenAI in public and develop actively in public repositories, so we expect this group to include early adopters of GenAI tools.
We call these users early adopters.
Some of their repositories contain AI-tool configuration files (ACFs), such as \texttt{.cursorrules} or \texttt{CLAUDE.md}, or list them in \texttt{.gitignore}.
We call these repositories \textit{repositories with ACFs} and the others \textit{repositories without ACFs}.
An ACF shows that the owner set up an AI coding tool for the repository.
Section~\ref{sec:threats} discusses the limits of these two labels.
Prior studies sample artifacts in which GenAI was used, such as self-admitted GenAI use~\cite{xiao2026selfadmitted}, ChatGPT conversations~\cite{xiao2024devgpt}, generated code committed to repositories~\cite{grewal2024chatgpt,siddiq2024quality}, pull requests created by agents~\cite{li2025rise,watanabe2026agentic}, and projects that build AI~\cite{yang2023github,cai2025demystifying}.
Their data therefore do not include the projects in which early adopters did not configure an AI tool.
We first select early adopters and then compare the repositories with ACFs and the repositories without ACFs that these early adopters own.}
We identified \tsuchida{531 early adopters} and compared \tsuchida{100} of their \tsuchida{repositories with ACFs} with \tsuchida{control repositories without ACFs, matched on primary language, commit count, contributor count, star count, and repository age}. We also \tsuchida{classified 370 issues from the two repository groups by GenAI relatedness, Type, and Outcome}.
To guide the empirical study, we pose two research questions (RQs).

\RqOne\ We divide this question into two parts: \RqOneOne\ compares \tsuchida{the distributions of software categories between the two repository groups}, and \RqOneTwo\ \tsuchida{compares six README structural characteristics between them: README character count, H1 header density, code-block density, image density, URL density, and license presence}.

\RqTwo\ We \tsuchida{classify each issue by GenAI relatedness, Type, and Outcome.
GenAI relatedness shows whether an issue concerns GenAI technology, and we call such issues \textit{GenAI-related issues}.
GenAI-related issues can come from implementing GenAI features, so we compare Type and Outcome with GenAI relatedness held fixed.
We call repositories that implement or depend on GenAI features \textit{GenAI-feature repositories}.
This term describes the function of the software and says nothing about how the developers wrote it.
We add this label to the \tsuchida{logistic mixed-effects model of RQ2 (Section~\ref{sec:rq2})} to separate GenAI features in the product from AI-tool use in development}.

\tsuchida{Our results show modest differences between the two repository groups in README structural characteristics, weak evidence of a difference in software categories, and no detected difference in their issues.
These README differences appear in four characteristics: README character count, code-block density, URL density, and license presence.
For license presence, 59\% of repositories with ACFs and 38\% of repositories without ACFs have a license.
The distribution of software categories differs between the two repository groups, but no single category differs after the Benjamini-Hochberg procedure, so we report this difference as weak evidence.
In the issues, GenAI-related issues are 64\% of the issues in repositories with ACFs and 44\% in repositories without ACFs, but this difference is not significant.
Once GenAI relatedness is held fixed, we detect no difference in Type or Outcome between the two repository groups.}

The remainder of this paper is organized as follows.
Section~\ref{sec:related} reviews related work.
Section~\ref{sec:data_preparation} provides an overview of data preparation.
\tsuchida{Section~\ref{sec:user_profile}} describes \tsuchida{early adopter} identification through user profile analysis\tsuchida{, Sections~\ref{sec:repo_selection} and~\ref{sec:issue_collection} describe the repository selection and matching and the issue collection, and Section~\ref{sec:ethics}} outlines our research ethics and compliance.
Section~\ref{sec:empirical} presents repository comparison and issue analysis for both research questions.
Section~\ref{sec:discussion} discusses implications.
Section~\ref{sec:threats} addresses threats to validity.
Section~\ref{sec:conclusion} concludes with future directions.
Our replication package is available at \url{https://doi.org/10.5281/zenodo.18302942}.

\section{Related Work}
\label{sec:related}

Research on GenAI in software engineering has grown rapidly since ChatGPT's release in November 2022. We organize related work into studies of GenAI capabilities, adoption and developer practice, repository-level evidence, issue-level challenges, and repository categorization.

\subsection{GenAI Capabilities and Generated Code Quality}

Prior work has examined GenAI as a new form of software engineering capability. Hassan et al.~\cite{hassan2026se3} proposed Software Engineering 3.0, envisioning AI systems as intelligent teammates collaborating throughout the development lifecycle. Large-scale studies of agent-generated code have characterized output quality: Li et al.~\cite{li2025rise} analyzed over 456,000 pull requests from five GenAI agents across 61,000 repositories, documenting challenges in acceptance rates and code quality, while Watanabe et al.~\cite{watanabe2026agentic} examined 567 agent-generated pull requests, reporting 83.8\% acceptance but 45.1\% requiring human revision before merging.

Systematic reviews have synthesized research on LLMs for software engineering. Hou et al.~\cite{hou2024large} analyzed 395 papers, categorizing 85 specific tasks across six core activities, with software maintenance accounting for 24.89\% of research. He et al.~\cite{he2025multiagent} reviewed LLM-based multi-agent systems, finding research publications increased from 1 paper in 2023 to 10 papers in 2024, indicating the domain is still in its formative period. These studies establish what GenAI systems can produce and where research attention is moving, but they do not show how \tsuchida{early adopters} organize \tsuchida{repositories with ACFs} or what \tsuchida{characteristics} appear in those repositories \tsuchida{and their issues}. Our work complements this literature by shifting the unit of analysis from model output to \tsuchida{the repositories of early adopters and the issues in these repositories}.

\subsection{GenAI Adoption and Developer Practice}

Empirical studies have documented how developers adopt and interact with GenAI tools. Xiao et al.~\cite{xiao2026selfadmitted} analyzed self-admitted GenAI usage across GitHub repositories, constructing a taxonomy of development tasks and finding that GenAI adoption does not lead to general increases in code churn. Giray et al.~\cite{giray2026genai} surveyed software engineering professionals and measured adoption in practice. Studies of developer interaction patterns further show how practitioners work with these tools: Xiao et al.~\cite{xiao2024devgpt} introduced DevGPT, a dataset of ChatGPT prompts linked to GitHub artifacts; Wang et al.~\cite{wang2024characterizing} characterized developer behavior in LLM-supported development; and Barke et al.~\cite{barke2023grounded} identified bimodal Copilot interaction patterns between acceleration and exploration.

Industry surveys and enterprise deployments provide complementary evidence about organizational use. GitHub's 2024 survey found widespread AI tool usage at work~\cite{github2024survey}. Liang et al.~\cite{liang2024large} and Sergeyuk et al.~\cite{sergeyuk2025using} reported both motivations and concerns, including keystroke reduction, output control, quality, and reliability. Enterprise studies such as Murali et al.'s CodeCompose deployment at Meta~\cite{murali2024ai} and Weisz et al.'s watsonx Code Assistant study~\cite{weisz2025examining} show how AI coding tools behave in organizational settings. These studies explain self-reported, observed, or enterprise-mediated adoption, while our study complements them by identifying \tsuchida{early} adopters through public GitHub signals and comparing their \tsuchida{repositories with ACFs and repositories without ACFs}.

\subsection{Repository-Level Evidence of GenAI Use}

Repository-level studies have examined where GenAI artifacts appear and how generated code behaves once committed. Yang et al.~\cite{yang2023github} analyzed issues from open-source \tsuchida{projects that build AI} and identified recurring problem categories. Grewal et al.~\cite{grewal2024chatgpt} studied ChatGPT-generated code snippets in GitHub repositories, finding that generated code is often retained and remains unchanged once added. Siddiq et al.~\cite{siddiq2024quality} analyzed ChatGPT-generated code from the DevGPT dataset and found quality issues including undefined variables and security vulnerabilities.

These studies provide important evidence about \tsuchida{projects that build AI} and generated code, but they focus on \tsuchida{these projects}, generated snippets, or artifact-level quality rather than comparing repositories created by \tsuchida{early adopters who state interest in GenAI on their public profiles}. \tsuchida{As Section~\ref{sec:introduction} notes, these studies sample only artifacts in which GenAI was used, so their data do not include the projects in which early adopters did not configure an AI tool.} Our work complements this line by comparing \tsuchida{repositories with ACFs} with \tsuchida{control repositories without ACFs} from the same \tsuchida{population of early adopters}, allowing us to examine repository categories and \tsuchida{README structural characteristics} associated with \tsuchida{ACFs}.

\subsection{GenAI-Related Issues and \tsuchida{Challenges in LLM Development}}

Research on GenAI-related issues has identified recurring challenges in LLM development and use. Chen et al.~\cite{chen2025llmchallenges} examined OpenAI developer forum questions and constructed a taxonomy of LLM developer challenges spanning prompts, APIs, and plugins. Cai et al.~\cite{cai2025demystifying} analyzed issues from \tsuchida{open-source projects that build LLMs} and found that Model Issue is the most common problem type.

These studies explain challenges in LLM ecosystems, but they do not analyze \tsuchida{the issues in the repositories of early adopters}. Our work complements them by \tsuchida{comparing the issues in repositories with ACFs with the issues in repositories without ACFs, with GenAI relatedness held fixed}.

\subsection{Repository Categorization}

Research on repository categorization provides methodological foundations for comparing repository types. Zanartu et al.~\cite{zanartu2022automatically} proposed machine learning approaches for automatically categorizing GitHub repositories by application domain, which we adopt in our RQ1 analysis.

This prior work supplies the classification method, while our contribution is to apply it to \tsuchida{the comparison of repositories with ACFs and repositories without ACFs}. By using repository categorization alongside \tsuchida{README structural characteristics} and issue classification, our study connects methodological work on repository domains with empirical evidence about \tsuchida{the repositories and issues of early adopters}.

\section{Data Preparation}
\label{sec:data_preparation}

\subsection{Overview of Data Collection}

We used GitHub's REST and GraphQL APIs to collect user-profile, repository, and issue datasets, summarized in Table~\ref{tab:data_summary}. The user-profile dataset identifies \tsuchida{early adopters}. The repository dataset supports RQ1 by matching repositories with \tsuchida{ACFs} to comparable \tsuchida{repositories without ACFs}. The issue dataset supports RQ2 by capturing issues created in \tsuchida{the two repository groups}. Data were collected from May 25 to August 14, 2025\tsuchida{, except for three of the matching variables, which we took on 15 October 2026 (Section~\ref{sec:threats})}.

\begin{table}[!htbp]
\centering
\caption{Summary of collected data}
\label{tab:data_summary}
\begin{tabular}{@{}llr@{}}
\toprule
\textbf{Analysis} & \textbf{Data} & \textbf{Count} \\
\midrule
Preliminary & GitHub User Profiles & \tsuchida{531} \\
\midrule
\multirow{2}{*}{RQ1} & \tsuchida{Repositories with ACFs} & \tsuchida{100} \\
 & \tsuchida{Repositories without ACFs} & \tsuchida{100} \\
\midrule
RQ2 & Issues & \tsuchida{370} \\
\bottomrule
\end{tabular}
\end{table}

\begin{table}[!htbp]
\centering
\footnotesize
\caption{\tsuchida{Counts at each step of the sample selection}}
\label{tab:funnel}
\begin{tabular}{@{}>{\raggedright\arraybackslash\leavevmode}p{0.82\linewidth}r@{}}
\toprule
\tsuchida{Step} & \tsuchida{Count} \\
\midrule
\multicolumn{2}{@{}l@{}}{\tsuchida{Users}} \\
\tsuchida{Search profiles for ``AI Agent'' and ``Vibe Coding''} & \tsuchida{1,084} \\
\tsuchida{Exclude duplicate accounts (found by both keywords)} & \tsuchida{1,083} \\
\tsuchida{Exclude early adopters with fewer than 10 public repositories} & \tsuchida{698} \\
\tsuchida{Exclude early adopters with fewer than 10 commits in the previous year} & \tsuchida{552} \\
\tsuchida{Exclude bots, organizations, and anti-AI accounts by manual inspection} & \tsuchida{531} \\
\midrule
\multicolumn{2}{@{}l@{}}{\tsuchida{Repositories}} \\
\tsuchida{All repositories owned by the remaining early adopters} & \tsuchida{30,326} \\
\tsuchida{Exclude duplicate repositories (collected for more than one early adopter)} & \tsuchida{30,312} \\
\tsuchida{Exclude repositories not updated within six months of the collection date} & \tsuchida{8,625} \\
\tsuchida{Exclude forks and repositories not owned by the early adopter (repositories of other users or organizations)} & \tsuchida{5,562} \\
\tsuchida{Exclude non-English repositories (including empty READMEs and READMEs too short to detect a language)} & \tsuchida{3,621} \\
\tsuchida{Exclude repositories that the GitHub API no longer returns in 2026 (deleted, renamed, or made private)} & \tsuchida{3,139} \\
\tsuchida{Exclude repositories whose rewritten history holds no commit before 8 July 2025, because the root file list on that date cannot be restored} & \tsuchida{3,130} \\
\tsuchida{Repositories with AI-tool configuration files} & \tsuchida{100} \\
\tsuchida{Repositories without AI-tool configuration files (candidates)} & \tsuchida{3,030} \\
\tsuchida{Assign one control with the same primary language to each repository with AI-tool configuration files} & \tsuchida{100 pairs} \\
\midrule
\multicolumn{2}{@{}l@{}}{\tsuchida{Issues (repositories with AI-tool configuration files / controls)}} \\
\tsuchida{All issues in the matched repositories (any author)} & \tsuchida{2,550 / 349} \\
\tsuchida{Exclude issues created before 30 November 2022 (the release of ChatGPT)} & \tsuchida{2,550 / 349} \\
\tsuchida{Exclude issues created after 14 August 2025} & \tsuchida{142 / 241} \\
\tsuchida{Exclude non-English issues} & \tsuchida{138 / 232} \\
\bottomrule
\end{tabular}
\end{table}

Because RQ1.1 compares repository types, we use the five-category taxonomy by Zanartu et al.~\cite{zanartu2022automatically}. Table~\ref{tab:repo_categories} summarizes these categories.

\begin{table*}[!htbp]
\centering
\caption{Repository Category Definitions~\cite{zanartu2022automatically}}
\label{tab:repo_categories}
\small
\begin{tabular}{@{}p{3.6cm}p{2.6cm}p{7cm}@{}}
\toprule
\textbf{Category} & \textbf{Examples} & \textbf{Definition} \\
\midrule
Application \& System Software & WordPress, Linux & Software that provides end-user functionalities or infrastructure services. \\
\addlinespace
Web Libraries and Frameworks & Bootstrap, Angular.js & Libraries and frameworks specifically designed to support web development. \\
\addlinespace
Non-web Libraries and Frameworks & Guava, Fresco & General-purpose libraries and frameworks used for non-web development contexts. \\
\addlinespace
Software Tools & Homebrew, Git & Tools that facilitate software engineering tasks, such as IDEs, compilers, and package managers. \\
\addlinespace
Documentation & Java Design Patterns & Repositories dedicated to informational content, including tutorials and code examples, rather than functional software. \\
\bottomrule
\end{tabular}
\end{table*}

\subsection{Identifying \tsuchida{Early Adopters}}
\label{sec:user_profile}

\tsuchida{Early adopters who state interest in GenAI on their public profiles are} useful for studying \tsuchida{the repositories and issues of early adopters of GenAI tools} because their repositories and issues \tsuchida{are public and record} real-world development.

We identified \tsuchida{early adopters} through a keyword search followed by activity and validity filters. The initial GitHub profile search used ``AI Agent'' \tsuchida{and} ``Vibe Coding''~\cite{karpathy2025vibecoding}, which \tsuchida{refer to} autonomous coding agents \tsuchida{and} natural-language-driven coding. We selected these keywords through team discussion. We did not search for tool names such as ``Copilot'' or ``Cursor'' because a sample seeded with specific tool names would over-represent users of those tools. The \tsuchida{two} phrases describe the development practice rather than a particular product, although they can miss \tsuchida{early adopters} who \tsuchida{state their interest} only through tool names, a limitation we discuss in Section~\ref{sec:threats}. This search returned \tsuchida{1,084} candidate profiles. \tsuchida{One account was found by both keywords, and removing the duplicate left 1,083.} To exclude brand-new accounts and developers with little activity, we removed profiles with fewer than 10 public repositories or fewer than 10 commits in the preceding year. The repository threshold removed \tsuchida{385} profiles, leaving \tsuchida{698}. The commit threshold removed a further \tsuchida{146}, leaving \tsuchida{552}.
One author then manually inspected the remaining profiles. Because GitHub also hosts organization accounts, which do not fit our focus on individual developers, we excluded \tsuchida{11} bot or organization accounts. The keyword search also captures developers who use these terms to express opposition, so we excluded 10 profiles signaling an anti-AI stance, such as bios reading ``vibe coding hater'' or ``hate vibe coding''. This process reduced the initial candidate set to the \tsuchida{early adopter} dataset summarized in Table~\ref{tab:data_summary}. Table~\ref{tab:funnel} reports the count remaining after each step, ending with \tsuchida{531} \tsuchida{early adopters}.

\tsuchida{To check whether the early adopters develop software with AI coding tools, we inspected a random sample of 50 of the 531 early adopters and found visible traces of AI coding tools for 32 of them (64\%). We drew the sample with a fixed random seed. Two coders independently inspected the non-fork repositories owned by each sampled early adopter for the four types of traces that Khosravani and Mockus~\cite{khosravani2026detecting} use: commits authored by AI agents, AI co-author trailers in commit messages, ACFs, and README statements about AI-tool use. An early adopter showed a trace when at least one of these repositories contained at least one of the four types. Cohen's kappa between the two coders was 0.77, with agreement on 45 of the 50 early adopters, and the coders resolved the disagreements by discussion. Because many forms of AI-tool use leave no trace in a repository, the share of 64\% is a lower bound.}

\subsection{\tsuchida{Early Adopter} Characteristics}

We characterized the \tsuchida{531} \tsuchida{early adopters} using account creation dates and professional-background heuristics. Account creation dates from GitHub's API approximate experience level. Bio, company, and README keywords such as ``research'' and ``researcher'' distinguish academic researchers from industry practitioners.
Out of \tsuchida{531} \tsuchida{early adopters}, \tsuchida{388} (73\%) users had established GitHub histories before ChatGPT's release, \tsuchida{so most early adopters are developers who used GitHub before GenAI tools became widely available}. The sample is also dominated by non-research practitioners\tsuchida{ (475 of the 531 early adopters, 89\%)}. The user-profile dataset includes user ID, account creation date, bio, follower count, repository count, profile README, organizations, and company affiliation.

\subsection{\tsuchida{Repository Selection and Matching}}
\label{sec:repo_selection}

\tsuchida{We selected 3,130 repositories from the 30,312 repositories that the 531 early adopters owned at the repository collection time, 8 July 2025 (Table~\ref{tab:funnel}). First, we kept the 8,625 repositories updated within six months before this time, following the activity filter that Codabux et al.~\cite{codabux2024teaching} recommend for selecting active repositories. Second, we excluded forks, because a fork keeps the history of the original repository and does not show the development activity of the early adopter. We also excluded repositories that the early adopter's own account does not own, to which the early adopter only contributes, and repositories owned by organization accounts, so that the remaining 5,562 repositories are those that the early adopters created themselves. Third, we kept the 3,621 repositories whose README text is in English and removed repositories with an empty README or a README too short to detect a language, so that the README analysis compares READMEs written in the same language.}

\tsuchida{The last two steps removed repositories whose data we could not retrieve for the matching or for detecting ACFs. Fourth, we excluded the repositories that the GitHub API no longer returns in 2026 because they were deleted, renamed, or made private, which left 3,139. We needed the repositories in 2026 because we took three of the matching variables on 15 October 2026 (Section~\ref{sec:threats}). Fifth, we excluded nine repositories whose history was rewritten and holds no commit before 8 July 2025, because their root file list on that date cannot be restored, which left 3,130.}

\tsuchida{The 3,130 repositories comprise 100 repositories with ACFs and 3,030 candidates without ACFs, based on the files that each repository had at the repository collection time. We detected ACFs at this time so that the group of each repository reflects its state at the same reference point as the other selection steps. Because the GitHub API returns the files of a repository at a past commit, we took the root file list and the .gitignore file from the last commit before 8 July 2025.} A repository was labeled as \tsuchida{a repository with ACFs} if its root directory contained configuration files associated with major GenAI agents or Model Context Protocol servers. Because some \tsuchida{early adopters} list these configuration files in .gitignore instead of committing them to the root directory, we also treated .gitignore entries that reference such configuration files as \tsuchida{ACFs}, since these files themselves are not committed. For each repository, we collected default branch, primary language, root files, .gitignore content, file count, directory count, commit count, \tsuchida{contributor count, star count, creation date,} issue counts by status, and README content\tsuchida{, taking commit count, contributor count, and primary language on 15 October 2026 (Section~\ref{sec:threats})}.

\tsuchida{We matched each of the 100 repositories with ACFs to one control repository among the 3,030 candidates on five variables, primary language, commit count, contributor count, star count, and repository age. Repository age and commit count approximate the maturity of a project, commit count and contributor count reflect its size~\cite{zhaoCI2017}, and star count and contributor count approximate its popularity and community size. The primary language had to be the same within each pair. Because the four numeric variables differ in unit and scale, we log-transformed and standardized them and gave them the same weight. The Hungarian algorithm~\cite{kuhn1955hungarian} then assigned the control repositories so that the total distance over the 100 pairs is the smallest, and every repository with ACFs received a control repository with the same primary language.}

\tsuchida{The matching reduced the differences between the two repository groups in all four numeric variables (Table~\ref{tab:smd}). We measured each difference as the standardized mean difference, the difference in means divided by the standard deviation. The before column compares the 100 repositories with ACFs with the 3,030 candidates, and the after column compares the 100 matched pairs. After matching, the largest absolute value over the four numeric variables is 0.081. The primary language is the same within each pair, so its difference after matching is 0.}

\begin{table}[!htbp]
\centering
\caption{\tsuchida{Standardized mean difference before and after matching}}
\label{tab:smd}
\begin{tabular}{@{}lrr@{}}
\toprule
\tsuchida{Variable} & \tsuchida{Before} & \tsuchida{After} \\
\midrule
\tsuchida{Commit count} & \tsuchida{+0.641} & \tsuchida{+0.078} \\
\tsuchida{Contributor count} & \tsuchida{+0.383} & \tsuchida{+0.081} \\
\tsuchida{Star count} & \tsuchida{+0.110} & \tsuchida{+0.064} \\
\tsuchida{Repository age} & \tsuchida{$-$0.547} & \tsuchida{$-$0.064} \\
\bottomrule
\end{tabular}
\end{table}

\subsection{\tsuchida{Issue Collection}}
\label{sec:issue_collection}

\tsuchida{We selected 370 issues from the two repository groups in two steps (Table~\ref{tab:funnel}). The repositories with ACFs had 2,550 issues and the repositories without ACFs had 349 issues. First, we excluded the issues created before 30 November 2022 or after 14 August 2025, which left 142 and 241 issues. We started the period at the release of ChatGPT so that developers had GenAI tools available when they wrote each issue, and we ended it on 14 August 2025, the date of our first issue collection, so that we observe the two repository groups over the same period. Second, we excluded non-English issues, which left 138 and 232 issues. As with the READMEs, we kept only English issues so that the two coders classify every issue in the same language with the same codebook. The 370 issues come from 26 repositories\tsuchida{, 13 in each group,} and \tsuchida{222} issue authors. \tsuchida{The other 174 matched repositories have no issue in this period.}} For each issue, we collected title, body text, creation date, closure status, and author information.

\subsection{Research Ethics and Compliance}
\label{sec:ethics}

We collected only public data through GitHub's official REST and GraphQL APIs, followed authenticated rate limits of 5,000 requests per hour, and inserted delays to avoid excessive load. We did not access private repositories or restricted content. Our analysis focused on public metadata, repository contents, commits and issue activity. In accordance with GitHub's policy for research using public information, this paper will be published as open access.

\section{Empirical Study}
\label{sec:empirical}

\tsuchida{This section reports the empirical evidence for the differences between the two repository groups in repository characteristics (RQ1) and in issue characteristics (RQ2).}

\subsection{Repository Comparison for RQ1}

\tsuchida{We compared the 100 pairs of a repository with ACFs and a control repository, selected and matched as described in Section~\ref{sec:repo_selection}, on repository category (RQ1.1) and on six README structural characteristics (RQ1.2).}

\tsuchida{For RQ1.1, we compared the distributions of software categories between the two repository groups. We applied the automated classification method by Zanartu et al.~\cite{zanartu2022automatically}, which uses repository metadata and textual features to classify repositories into the categories shown in Table~\ref{tab:repo_categories}. We then used the Stuart-Maxwell test, which takes each matched pair as the unit, to compare category distributions between repositories with ACFs and repositories without ACFs. This is a single test, so we did not adjust its p value. The 100 repositories with ACFs belong to 61 owners, and the 100 repositories without ACFs belong to 75 owners. Of these owners, 21 and 19 own two or more repositories, so repositories of the same owner may form clusters. We therefore also fitted a mixed-effects model for each of the five categories with the repository owner as a random intercept, and adjusted the five p values with the Benjamini-Hochberg procedure.}

For RQ1.2, we compared \tsuchida{six README structural characteristics of the two repository groups}. We measured \tsuchida{six document features: README character count}, H1 header density, code-block density, image density, URL density, and license presence. Because README size varies widely across repositories, we normalized the H1 header, code-block, image, and URL counts to a rate per 1,000 characters of README text so that repositories with different README lengths are comparable. We kept the total character count as a raw value and recorded license presence as a binary indicator for each repository, reported as the percentage of repositories that include a license. License presence was measured with a keyword-based binary classifier covering general license terms, specific license names, versioned licenses, license-file references, Software Package Data Exchange (SPDX) identifiers, and legal terminology. The full keyword list is available in our replication package.

\tsuchida{Because each repository with ACFs is paired with one control repository, we assessed statistical significance using paired tests: the Wilcoxon signed-rank test for the five continuous metrics and McNemar's test with the continuity correction for license presence. The six metrics come from the same document and are correlated, so we adjusted the six p values with the Benjamini-Hochberg procedure. We report Cliff's $\delta$ as the effect size for the five continuous metrics~\cite{cliff1993dominance,romano2006exploring} and the difference in shares for license presence. We computed a 95\% confidence interval for Cliff's $\delta$ from 2,000 bootstrap resamples of the pairs, and a Wald 95\% confidence interval for matched pairs for the difference in shares. As for RQ1.1, we also fitted a mixed-effects model with the repository owner as a random intercept. The five continuous metrics are strongly skewed, so we fitted a linear mixed-effects model to $\log(x+1)$ of each of them. For license presence, we fitted a logistic mixed-effects model.}

\paragraph{Findings}
\begin{figure*}[!t]
\centering
\includegraphics[width=0.8\textwidth]{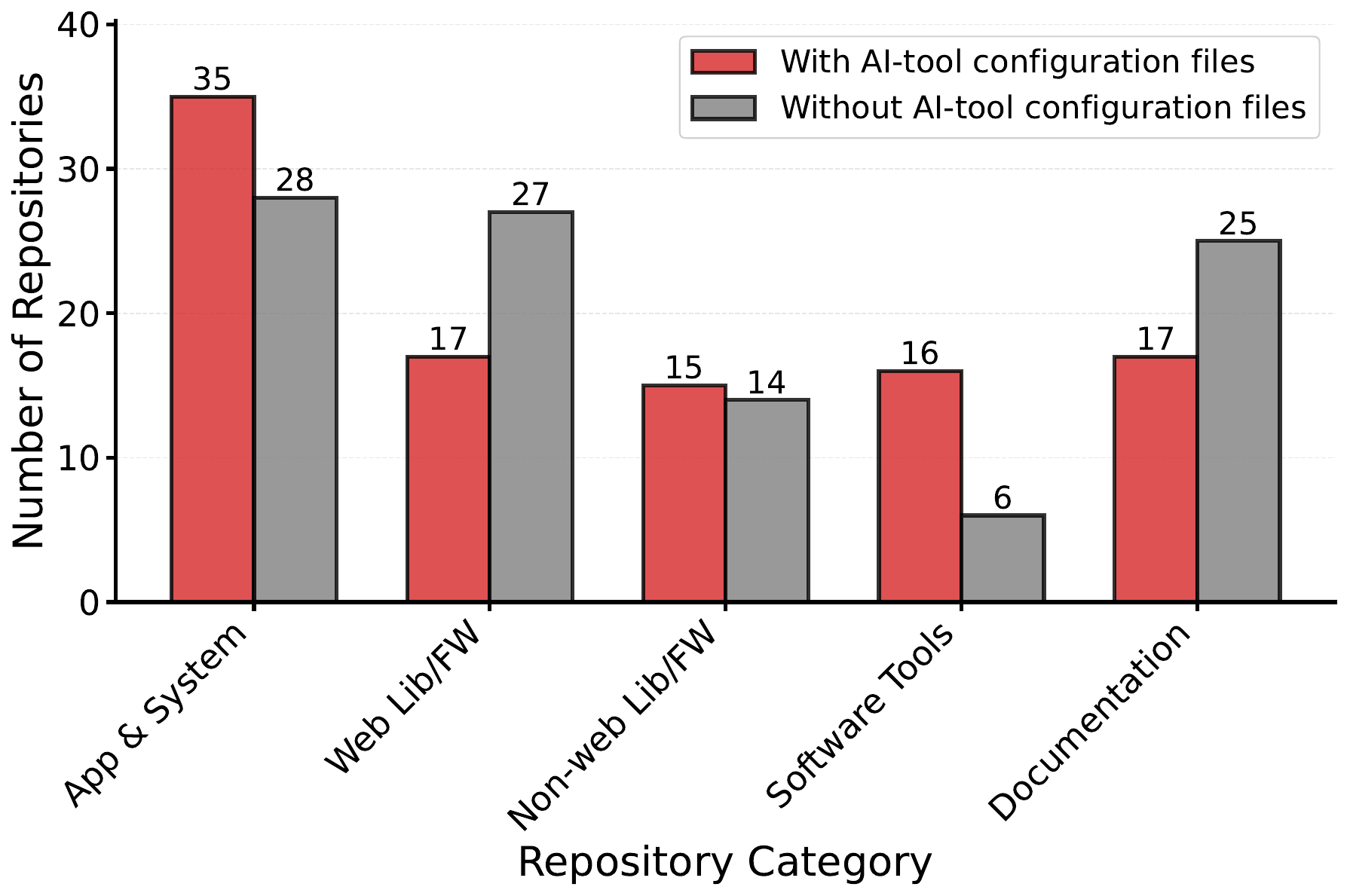}
\caption{Repository category distribution comparing \tsuchida{100 repositories with ACFs and 100 repositories without ACFs}}
\label{fig:repo_categories}
\end{figure*}

\tsuchida{%
The category distributions of the two repository groups show weak evidence of a difference.
Figure~\ref{fig:repo_categories} presents the functional category distribution comparing the two repository groups. Repositories with ACFs were most frequently categorized as \textit{Application \& System Software}. In contrast, the most frequent categories among repositories without ACFs were \textit{Application \& System Software}, \textit{Web Libraries and Frameworks}, and \textit{Documentation}. The Stuart-Maxwell test detected a difference in category distribution, as shown in Table~\ref{tab:rq1_stats}. In the mixed-effects models, no category had an adjusted p value below 0.05.
The \textit{Software Tools} category was numerically higher among repositories with ACFs, and the \textit{Web Libraries and Frameworks} and \textit{Documentation} categories were numerically higher among repositories without ACFs, but the differences were not significant.}

\begin{figure*}[!t]
\centering
\includegraphics[width=0.8\textwidth]{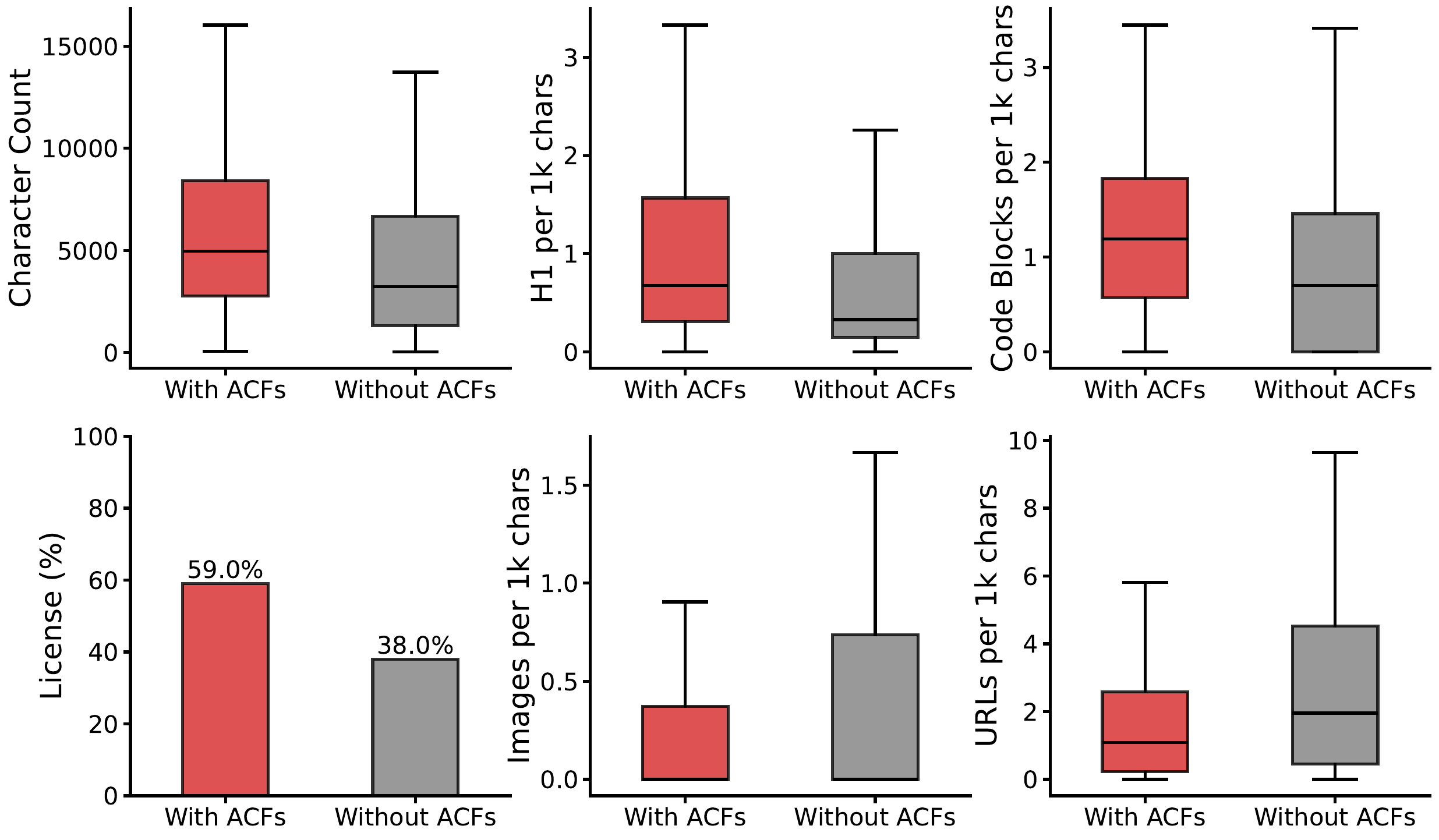}
\caption{\tsuchida{README structural characteristics of the two repository groups} across six metrics: Character Count, H1 per 1k chars, Code Blocks per 1k chars, License percentage, Images per 1k chars, and URLs per 1k chars}
\label{fig:readme_analysis}
\end{figure*}

\tsuchida{The two repository groups differ in four of the six README metrics. Figure~\ref{fig:readme_analysis} presents the README comparison, and Table~\ref{tab:rq1_stats} summarizes the statistical tests. Repositories with ACFs had significantly longer README files, with more code blocks per 1k characters. A longer README may also reflect long text that was cheaply generated. In contrast, repositories without ACFs contained more URLs per 1k characters, although we did not measure whether these links work. A license was present more often in repositories with ACFs (59\%) than in repositories without ACFs (38\%). H1 header density and image density showed no significant differences after adjustment. All three significant values of $\delta$ are small, and the mixed-effects models gave the same four differences.}

\begin{table}[!htbp]
\centering
\caption{Statistical test results for RQ1}
\label{tab:rq1_stats}
\resizebox{\columnwidth}{!}{%
\begin{tabular}{llrrr}
\toprule
Metric & Test & Statistic & \tsuchida{Adjusted p} & Effect Size\tsuchida{ [95\% CI]}\\
\midrule
Category Distribution & \tsuchida{Stuart-Maxwell} & \tsuchida{10.276} & \tsuchida{0.036\textsuperscript{a}} & --\\
\midrule
Character Count & \tsuchida{Wilcoxon signed-rank} & \tsuchida{1,814.0} & \tsuchida{0.032} & \tsuchida{0.237 [0.086, 0.373]} (small)\\
H1 per 1k chars & \tsuchida{Wilcoxon signed-rank} & \tsuchida{1,924.0} & \tsuchida{0.065} & \tsuchida{0.212 [0.076, 0.348]} (small)\\
Code Blocks per 1k chars & \tsuchida{Wilcoxon signed-rank} & \tsuchida{1,527.0} & \tsuchida{0.016} & \tsuchida{0.237 [0.090, 0.387]} (small)\\
URLs per 1k chars & \tsuchida{Wilcoxon signed-rank} & \tsuchida{1,465.0} & \tsuchida{0.010} & \tsuchida{$-$0.187 [$-$0.324, $-$0.038]} (small)\\
Images per 1k chars & \tsuchida{Wilcoxon signed-rank} & \tsuchida{627.0} & \tsuchida{0.077} & \tsuchida{$-$0.087 [$-$0.216, 0.044]} (negligible)\\
License Presence & \tsuchida{McNemar} & \tsuchida{7.547} & \tsuchida{0.016} & \tsuchida{0.210 [0.073, 0.347]}\\
\bottomrule
\multicolumn{5}{l}{\footnotesize \tsuchida{\textsuperscript{a}A single test, so the p value is not adjusted.}}\\
\multicolumn{5}{l}{\footnotesize Effect Size = Cliff's $\delta$\tsuchida{, except for license presence, for which the effect size is the difference in shares.}}\\
\end{tabular}%
}
\end{table}

\begin{tcolorbox}[colback=white, colframe=summaryheader, coltitle=white, colbacktitle=summaryheader, title=\textbf{RQ1 Summary}, fonttitle=\bfseries\large, boxrule=1.5pt, arc=3pt]
\tsuchida{Repositories with ACFs have longer READMEs, more code blocks per 1k characters, fewer URLs per 1k characters, and a license more often than repositories without ACFs. The category distributions of the two repository groups show weak evidence of a difference.}
\end{tcolorbox}

\subsection{Issue Analysis for RQ2}
\label{sec:rq2}

\tsuchida{We compared the 370 issues from the two repository groups, collected as described in Section~\ref{sec:issue_collection}, on GenAI relatedness, Type, and Outcome.}

\tsuchida{We labeled each issue on three dimensions: GenAI relatedness, Type, and Outcome.} To distinguish \tsuchida{issues that concern GenAI technology from other issues}, we labeled each issue as GenAI-related or \tsuchida{Non-GenAI-related} according to whether the issue itself concerned GenAI technology. The \tsuchida{second dimension, Type,} categorizes issues as Bug, Enhancement, \tsuchida{Question,} or Other, following the Natural Language-Based Software Engineering (NLBSE) '22 Tool Competition taxonomy~\cite{kallis2022nlbse}. The \tsuchida{third dimension}, Outcome, categorizes the affected area: External, Internal, UI/UX, Developer Experience, Environment, Documentation\tsuchida{, or Unknown. We added Unknown because a question that requests no concrete task cannot be given an Outcome value}. Table~\ref{tab:classification_rules} defines these Outcome \tsuchida{values with representative examples}. Some issues could reasonably fit more than one Outcome \tsuchida{value}. Using a single label is a simplification, and we labeled each issue by the area it affected most.

\tsuchida{Two coders} classified all issues\tsuchida{ with a codebook that defines each value of the three dimensions (Tables~\ref{tab:codebook_genai_type} and~\ref{tab:classification_rules}). After a calibration round on 20 issues, which was used to refine the codebook, the two coders independently classified 80 additional issues. One coder used an LLM to draft initial labels from the codebook and then reviewed and corrected every label manually before comparing the labels with those of the other coder. We drew the 80 issues by simple random sampling from the issues of the two repository groups, and the coders saw the repository name, the title, and the body of each issue but not its group. Cohen's kappa was 0.750 for GenAI relatedness, 0.783 for Type, and 0.785 for Outcome. The two coders disagreed on 10 issues for GenAI relatedness, 12 for Type, and 12 for Outcome, 31 issues in total, and the two coders discussed each of these disagreements until they agreed. The remaining 270 issues were split between the two coders and classified individually, and the same coder used the same procedure for the issues that this coder classified. The final results are based on the consensus labels and the individual labels. All labels and the codebook are available in our replication package.}

\tsuchida{The issues are concentrated in a few repositories and authors (Table~\ref{tab:concentration}), so we used the repository as the statistical unit.}

\begin{table}[!htbp]
\centering
\caption{\tsuchida{Concentration of issues by author and by repository}}
\label{tab:concentration}
\resizebox{\columnwidth}{!}{%
\begin{tabular}{lrrr}
\toprule
 & \tsuchida{All} & \tsuchida{With ACFs} & \tsuchida{Without ACFs}\\
\midrule
\tsuchida{Issues} & \tsuchida{370} & \tsuchida{138} & \tsuchida{232}\\
\tsuchida{Repositories} & \tsuchida{26} & \tsuchida{13} & \tsuchida{13}\\
\tsuchida{Top author} & \tsuchida{60 (16.2\%)} & \tsuchida{11 (8.0\%)} & \tsuchida{60 (25.9\%)}\\
\tsuchida{Top repository} & \tsuchida{89 (24.1\%)} & \tsuchida{44 (31.9\%)} & \tsuchida{89 (38.4\%)}\\
\bottomrule
\end{tabular}%
}
\end{table}

\tsuchida{We compared Type and Outcome between the two repository groups with logistic mixed-effects models, holding GenAI relatedness fixed, because GenAI-related issues can come from implementing GenAI features. For each value of Type and Outcome, we modeled whether an issue has that value, with whether its repository has ACFs and its GenAI relatedness as covariates and the repository as a random intercept, and we adjusted the 11 p values with the Benjamini-Hochberg procedure. We report the effect of ACFs as an odds ratio with repositories without ACFs as the reference. With a model of the same form that has GenAI relatedness as the dependent variable, we tested whether GenAI-related issues are more common in repositories with ACFs. This is a single test, so we did not adjust its p value.}

\tsuchida{To separate GenAI features in the product from AI-tool use in development, two coders classified each of the 26 repositories as a GenAI-feature repository or not from its README. Cohen's kappa was 0.92. Of the 26 repositories, 16 are GenAI-feature repositories, 6 with ACFs and 10 without ACFs. We added this label as a further covariate to the model for each of the 11 values.}

\paragraph{Findings}
\begin{figure*}[t]
\centering
\includegraphics[width=0.9\textwidth]{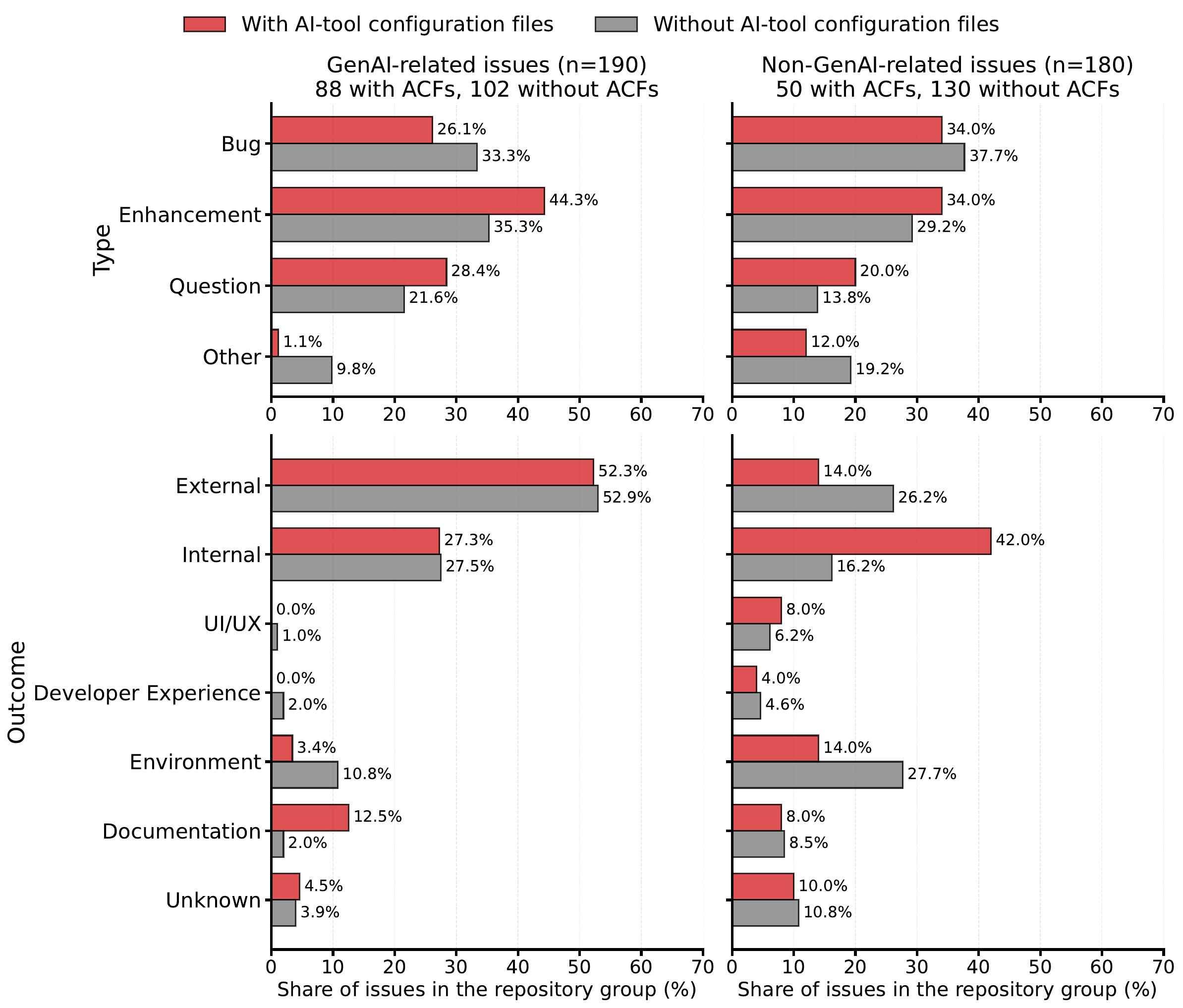}
\caption{\tsuchida{Shares of Type and Outcome in the issues of the two repository groups, for the 190 GenAI-related issues and the 180 Non-GenAI-related issues. Each percentage is the share of issues in one repository group within GenAI-related or Non-GenAI-related issues.}}
\label{fig:issue_taxonomy}
\end{figure*}

\begin{table*}[!htbp]
\centering
\begin{threeparttable}
\caption{\tsuchida{Codebook for GenAI relatedness and Type}}
\label{tab:codebook_genai_type}
\scriptsize
\begin{tabular}{@{}>{\raggedright\arraybackslash\leavevmode}p{0.15\linewidth}>{\raggedright\arraybackslash\leavevmode}p{0.81\linewidth}@{}}
\toprule
\textbf{\tsuchida{Value}} & \textbf{Description} \\
\midrule
\multicolumn{2}{@{}l}{\tsuchida{\textit{GenAI relatedness}}} \\
\tsuchida{GenAI} & \tsuchida{The issue itself concerns GenAI technology, such as LLMs, AI agents, MCP, prompts, embeddings, model output, or AI-related tools such as n8n, Crawl4AI, and Ollama.} \\
\tsuchida{Non-GenAI} & \tsuchida{The issue concerns general software concerns unrelated to GenAI, such as ports, Docker, or SSL, even when it occurs in a GenAI project.} \\
\midrule
\multicolumn{2}{@{}l}{\tsuchida{\textit{Type}}} \\
\tsuchida{Bug} & \tsuchida{The issue reports behavior that does not work as intended, such as an error, crash, or incorrect output, including cases where the writer asks about an error while being confident that it is a defect in the project.} \\
\tsuchida{Enhancement} & \tsuchida{The issue requests new or changed functionality, including the implementation of a missing feature.} \\
\tsuchida{Question} & \tsuchida{The issue asks for information or clarification without requesting a concrete change, including cases where the writer asks about an error without being sure whether it is a defect.} \\
\tsuchida{Other} & \tsuchida{The issue fits none of the above, such as notes, design memos, or test issues.} \\
\bottomrule
\end{tabular}
\end{threeparttable}
\end{table*}

\begin{table*}[!htbp]
\centering
\begin{threeparttable}
\caption{\tsuchida{Codebook for Outcome, with representative examples}}
\label{tab:classification_rules}
\scriptsize
\begin{tabular}{@{}>{\raggedright\arraybackslash\leavevmode}p{0.13\linewidth}>{\raggedright\arraybackslash\leavevmode}p{0.46\linewidth}>{\raggedright\arraybackslash\leavevmode}p{0.15\linewidth}>{\raggedright\arraybackslash\leavevmode}p{0.15\linewidth}@{}}
\toprule
\textbf{\tsuchida{Value}} & \textbf{Description} & \textbf{GenAI\tsuchida{-related} Example} & \textbf{Non-GenAI\tsuchida{-related} Example} \\
\midrule
\multicolumn{4}{@{}l}{\tsuchida{\textit{Outcome}}} \\
External & \tsuchida{The issue concerns how the project interacts with another service or tool, such as an LLM provider API, an MCP server, an observability service, a BaaS, or a bundled service, including changes to the project's own code for connecting to such a service and integration errors whose root cause lies in the project's own configuration.} & \tsuchida{Error when using o4-mini}\tnote{1} (Bug) & \tsuchida{Cannot import into LRC}\tnote{2} (Bug) \\
Internal & \tsuchida{The issue concerns the project's own code logic, such as data handling, control flow, or error handling, where the problem itself does not involve interaction with another service or tool, even if one is mentioned.} & \tsuchida{Need config for crawling}\tnote{3} (Enhancement) & \tsuchida{Support lower case keywords}\tnote{4} (Enhancement) \\
UI/UX & \tsuchida{The issue changes what end users see or interact with, such as screens, layout, or displayed text.} & \tsuchida{Easy change of model to avoid a memory error}\tnote{5} (Enhancement) & REPL unable to scroll vertically\tnote{6} (Bug) \\
Developer Experience & \tsuchida{The issue affects only the developers' own work, such as tooling, testing, formatting, or debugging setup.} & \tsuchida{Add list of test cases of prompts to main repo}\tnote{7} (Enhancement) & \tsuchida{cargo fmt -{}-all -{}-check}\tnote{8} (Bug) \\
Environment & \tsuchida{The issue concerns the development or runtime foundation itself, such as installation, configuration, dependencies, containers, setup, or the operating system, without involving interaction with another service or tool.} & \tsuchida{Use MCP with uvx}\tnote{9} (Enhancement) & \tsuchida{Apple Silicon support}\tnote{10} (Bug) \\
Documentation & \tsuchida{The issue's primary deliverable is documentation or a tutorial for users in any format, such as a README, specification, code example, or video.} & \tsuchida{Sample PLANNING.md \& TASK.md}\tnote{11} (Enhancement) & \tsuchida{Mention Git version}\tnote{12} (Enhancement) \\
\tsuchida{Unknown} & \tsuchida{The issue does not contain enough information to determine the affected area, such as a question with no clear task.} & \tsuchida{google/adk-python alignment?}\tnote{13} \tsuchida{(Question)} & \tsuchida{How do I use Lightroom with it?}\tnote{14} \tsuchida{(Question)} \\
\bottomrule
\end{tabular}
\begin{tablenotes}
\scriptsize
\item[1] \tsuchida{\url{https://github.com/coleam00/mcp-mem0/issues/2}}
\item[2] \tsuchida{\url{https://github.com/LYL1015/JarvisArt/issues/10}}
\item[3] \tsuchida{\url{https://github.com/coleam00/mcp-crawl4ai-rag/issues/17}}
\item[4] \tsuchida{\url{https://github.com/iankressin/eql/issues/31}}
\item[5] \tsuchida{\url{https://github.com/coleam00/local-ai-packaged/issues/57}}
\item[6] \url{https://github.com/iankressin/eql/issues/59}
\item[7] \tsuchida{\url{https://github.com/zk1tty/rebrowse-app/issues/1}}
\item[8] \tsuchida{\url{https://github.com/tuaregsand/Guardian-AA-A-Zero-Knowledge-AI-Wallet-Copilot/issues/3}}
\item[9] \tsuchida{\url{https://github.com/coleam00/local-ai-packaged/issues/37}}
\item[10] \tsuchida{\url{https://github.com/coleam00/local-ai-packaged/issues/121}}
\item[11] \tsuchida{\url{https://github.com/coleam00/context-engineering-intro/issues/1}}
\item[12] \tsuchida{\url{https://github.com/coleam00/local-ai-packaged/issues/70}}
\item[13] \tsuchida{\url{https://github.com/pontus-devoteam/adk-typescript/issues/1}}
\item[14] \tsuchida{\url{https://github.com/LYL1015/JarvisArt/issues/9}}
\end{tablenotes}
\end{threeparttable}
\end{table*}

\tsuchida{GenAI-related issues are more common in repositories with ACFs, but this difference is not significant. GenAI-related issues are 64\% of the issues in repositories with ACFs and 44\% of the issues in repositories without ACFs. The logistic mixed-effects model did not detect this difference (Table~\ref{tab:rq2_or}), because the 26 repositories differ more from each other than the two repository groups do.}

\tsuchida{Once GenAI relatedness is held fixed, we detect no difference in Type or Outcome between the two repository groups. Figure~\ref{fig:issue_taxonomy} shows the Type and Outcome distributions, and Table~\ref{tab:rq2_or} shows the odds ratios. No value has an adjusted p value below 0.05, and every confidence interval includes 1. Enhancement is more common in repositories with ACFs, and Bug is more common in repositories without ACFs, but neither difference is significant. Environment shows the largest difference in shares. For Other, the shares and the odds ratio differ in direction, because 34 of the 35 Other issues in repositories without ACFs come from two of the 13 repositories and the random intercept for the repository accounts for them.}

\tsuchida{External issues are associated with GenAI relatedness and show no detectable association with ACFs. External is 38\% of the issues in both repository groups. In the same model, GenAI-related issues concern External significantly more often than Non-GenAI-related issues (odds ratio 3.89, 95\% CI 2.29 to 6.62, adjusted p $<$ 0.001). Adding the GenAI-feature repository label to the models for all 11 values does not change these results. With the label in the models, the adjusted p value is 0.92 or higher for ACFs and 0.16 or higher for the label itself for every value.}

\begin{table}[!htbp]
\centering
\caption{\tsuchida{Shares and odds ratios of GenAI relatedness, Type, and Outcome in repositories with ACFs and repositories without ACFs}}
\label{tab:rq2_or}
\resizebox{\columnwidth}{!}{%
\begin{tabular}{lrrrr}
\toprule
\tsuchida{Value} & \tsuchida{With (\%)} & \tsuchida{Without (\%)} & \tsuchida{Odds ratio (95\% CI)} & \tsuchida{Adjusted p}\\
\midrule
\multicolumn{5}{@{}l}{\tsuchida{GenAI relatedness}}\\
\tsuchida{GenAI-related} & \tsuchida{64} & \tsuchida{44} & \tsuchida{1.14 (0.27--4.79)} & \tsuchida{0.86\textsuperscript{a}}\\
\midrule
\multicolumn{5}{@{}l}{\tsuchida{Type}}\\
\tsuchida{Bug} & \tsuchida{29} & \tsuchida{36} & \tsuchida{0.52 (0.15--1.75)} & \tsuchida{0.782}\\
\tsuchida{Enhancement} & \tsuchida{41} & \tsuchida{32} & \tsuchida{1.88 (0.67--5.28)} & \tsuchida{0.782}\\
\tsuchida{Question} & \tsuchida{25} & \tsuchida{17} & \tsuchida{1.10 (0.32--3.73)} & \tsuchida{0.882}\\
\tsuchida{Other} & \tsuchida{5} & \tsuchida{15} & \tsuchida{1.43 (0.05--41.57)} & \tsuchida{0.882}\\
\midrule
\multicolumn{5}{@{}l}{\tsuchida{Outcome}}\\
\tsuchida{Developer Experience} & \tsuchida{1} & \tsuchida{3} & \tsuchida{0.53 (0.11--2.57)} & \tsuchida{0.782}\\
\tsuchida{Documentation} & \tsuchida{11} & \tsuchida{6} & \tsuchida{2.18 (0.39--12.20)} & \tsuchida{0.782}\\
\tsuchida{Environment} & \tsuchida{7} & \tsuchida{20} & \tsuchida{0.35 (0.07--1.68)} & \tsuchida{0.782}\\
\tsuchida{External} & \tsuchida{38} & \tsuchida{38} & \tsuchida{0.65 (0.26--1.61)} & \tsuchida{0.782}\\
\tsuchida{Internal} & \tsuchida{33} & \tsuchida{21} & \tsuchida{1.44 (0.49--4.26)} & \tsuchida{0.788}\\
\tsuchida{UI/UX} & \tsuchida{3} & \tsuchida{4} & \tsuchida{0.46 (0.03--6.88)} & \tsuchida{0.788}\\
\tsuchida{Unknown} & \tsuchida{7} & \tsuchida{8} & \tsuchida{1.18 (0.30--4.70)} & \tsuchida{0.882}\\
\bottomrule
\multicolumn{5}{l}{\footnotesize \tsuchida{\textsuperscript{a}A single test, so the p value is not adjusted.}}\\
\end{tabular}%
}
\end{table}

\tsuchida{Within each Outcome value, GenAI-related and Non-GenAI-related issues concern different problems, as the examples in Table~\ref{tab:classification_rules} show.}

\tsuchida{\textit{External issues} differ in the kind of service that the project interacts with. GenAI-related External issues primarily involve GenAI API failures, such as an error returned by the OpenAI API when a memory tool sends a parameter that the o4-mini model does not support. Non-GenAI-related External issues involve integrations with other software, such as an error when a generated preset is imported into Adobe Lightroom Classic. This contrast is consistent with the architecture of GenAI applications, which often depend on external GenAI provider APIs for core functionality.}

\tsuchida{\textit{Internal issues} differ in the kind of system logic they address. GenAI-related Internal issues focus on organizing AI-specific behavior, such as limiting the pages that a crawler collects for a retrieval-augmented generation pipeline. Non-GenAI-related Internal issues address more conventional program logic, such as accepting lower-case keywords in a query language.}

\tsuchida{\textit{User Interface (UI/UX) issues} are more prominent among Non-GenAI-related issues, but the one GenAI-related UI/UX issue concerns a distinct interface problem. Its author asks for an easy way to change the model when the default model needs more memory than the user's machine has. Non-GenAI-related UI/UX issues address general interface problems, such as a terminal REPL that cannot scroll to show long query results.}

\tsuchida{\textit{Developer Experience issues} are less frequent in both GenAI-related and Non-GenAI-related issues, but they point to different development concerns. GenAI-related cases concern testing the behavior of an agent, such as adding a list of prompt test cases to the repository. Non-GenAI-related cases focus on general workflow tools, such as fixing a code formatter check that fails in continuous integration.}

\tsuchida{\textit{Environment issues} are less common among GenAI-related issues (7.4\%) than among Non-GenAI-related issues (23.9\%). GenAI-related Environment issues involve AI tool setup, such as running an MCP server with the uvx package runner. Non-GenAI-related Environment issues address traditional deployment problems, such as container image pulls that fail on Apple Silicon. The lower frequency in GenAI-related issues is consistent with GenAI applications that send AI processing to external APIs, which leaves less to set up in the local environment.}

\tsuchida{\textit{Documentation issues} were rare in both GenAI-related and Non-GenAI-related issues. In the GenAI-related example, the author requests sample \texttt{PLANNING.md} and \texttt{TASK.md} files that the \texttt{CLAUDE.md} of the repository refers to, while in the Non-GenAI-related example, the author asks for the README to mention the required Git version. Both requests ask for documentation for users, and the GenAI-related request concerns agent configuration.}

\tsuchida{The RQ2 comparison rests on 26 repositories, so it can detect only large differences, and several confidence intervals are wide. We therefore do not read these results as evidence that the two repository groups are the same.}

\begin{tcolorbox}[colback=white, colframe=summaryheader, coltitle=white, colbacktitle=summaryheader, title=\textbf{RQ2 Summary}, fonttitle=\bfseries\large, boxrule=1.5pt, arc=3pt]
\tsuchida{Once GenAI relatedness is held fixed, we detect no difference in Type or Outcome between repositories with ACFs and repositories without ACFs. External issues are associated with GenAI relatedness, and we detect no association with ACFs.}
\end{tcolorbox}

\section{Implications}
\label{sec:discussion}

From these findings, we draw practical guidance for developers and research opportunities for the software engineering community.

\subsection{For Developers on GitHub}

\textit{Developers should explicitly document external dependencies when \tsuchida{building GenAI features}.} \tsuchida{GenAI features often depend on remote provider APIs. Developers should therefore write the constraints of these APIs in their documentation, such as READMEs, including rate limits, authentication, model availability, and output formats. In RQ2, GenAI-related issues concern External more often than Non-GenAI-related issues (Section~\ref{sec:rq2}).} \tsuchida{Chen et al. also report} that external tool integration is a main challenge in GenAI application development~\cite{chen2025llmchallenges}. \tsuchida{In RQ2, External issues are associated with GenAI relatedness, and we detect no association between External issues and ACFs, so we base this advice on whether a project builds GenAI features.}

\tsuchida{\textit{Developers have always needed to check their READMEs and licenses, and developers who let AI coding tools write these files should check them more closely.} Repositories with ACFs had longer READMEs, more code blocks per 1k characters, fewer URLs per 1k characters, and a license more often (Section~\ref{sec:empirical}). When an AI coding tool writes a license, the developer should check that it is the intended one. A README is the first document that users and new contributors read, so a human should read it at least as carefully as the code, although a longer README takes more time to read. When an AI coding tool writes a README, developers may also choose the URLs that users need, such as links to the documentation of dependencies, and add them by hand. We did not observe who wrote these files, so these differences do not necessarily mean that AI coding tools wrote them.}

\subsection{For SE Researchers}

\tsuchida{\textit{Studies of GenAI-related issues should separate GenAI features in the product from AI-tool use in development.} In RQ2, External issues are associated with GenAI relatedness, and we detect no association with ACFs, even after we add the GenAI-feature repository label (Section~\ref{sec:rq2}). A study that collects issues only from repositories with ACFs cannot separate these two sources of External issues. Our comparison rests on only 26 repositories, and several confidence intervals are wide. Issues in the same repository are correlated and add less information than issues from new repositories. Later studies that compare the two repository groups should therefore plan the number of repositories with a power analysis.}

\tsuchida{\textit{Our data, collected from May to August 2025, describe how early adopters develop software in Software Engineering 2.0~\cite{hassan2026se3}, and later studies can use these data as a point of comparison.} In Software Engineering 2.0, most software development falls between the assistant mode and the agent mode of AI support (Section~\ref{sec:introduction}). Our data show the repositories for which early adopters set up AI coding tools, how these repositories differ from repositories without ACFs in README structural characteristics, and which issues arise in the two repository groups. When development reaches Software Engineering 3.0, later studies can collect the same data with the README metrics and the codebook in our replication package. Our data thus enable these studies to trace how developers' use of AI coding tools and the issues in their repositories change from Software Engineering 2.0 to Software Engineering 3.0.}

\section{Threats to Validity}
\label{sec:threats}

This section discusses limitations related to matching, classification, sampling, and the observable signals used to identify \tsuchida{early adopters and ACFs}.

\paragraph{Internal Validity}

\tsuchida{Two factors related to the matching} may affect the internal validity of our results. First, our repository matching relies \tsuchida{on five variables (Section~\ref{sec:repo_selection})}, which cannot account for confounding factors such as repository domain. \tsuchida{We did not match on the domain, because the type of software is the outcome of RQ1.1, and matching on it would remove the difference that RQ1.1 tests.} \fixme{\tsuchida{Second, the timing of the matching variables may also affect the matching. We took three of the matching variables, commit count, contributor count, and primary language, on 15 October 2026, about 15 months after we selected the repositories on 8 July 2025, and we could not restore any of the three for the selection date. The other two matching variables, star count and repository age, come from the data collected on 8 July 2025. The GitHub API does not return the contributor count or the primary language for a past date. We tried to restore the commit count on 8 July 2025 from the commit history, but for the 51 repositories in the analysis whose commit count we had recorded on that date, the restored value matched the record for only 39; a force push or a rebase, which rewrites the history, is one possible cause. If the two repository groups changed in different ways after the selection, pairs that are similar in October 2026 may have differed at the time of selection. We therefore used the values from 15 October 2026 for all three variables and both repository groups, so that the two groups are compared at the same point in time.}}

\tsuchida{Two factors related to the issue classification may also affect the internal validity.} First, issue classification was performed manually\tsuchida{, and two coders classified only 100 of the 370 issues. Cohen's kappa on the 80 issues that the two coders classified independently was 0.750 for GenAI relatedness, 0.783 for Type, and 0.785 for Outcome, but a single coder classified each of the other 270 issues, so their labels lack an inter-rater agreement measure}. To address potential subjectivity, we adopted an explicit classification framework (\tsuchida{Tables~\ref{tab:codebook_genai_type} and~\ref{tab:classification_rules}}) and resolved \tsuchida{the disagreements between the two coders through discussion}. \tsuchida{Second}, our classification assigns each issue to a single \tsuchida{Outcome value} (e.g., External, Internal, UI/UX), although some issues may span multiple \tsuchida{values}. While \tsuchida{a single value} provides a clear distinction for analysis, it may oversimplify complex issues. Future work could explore multi-label classification to capture issues affecting multiple system aspects simultaneously.

\paragraph{External Validity}

Our sample represents publicly self-identified early adopters. The keyword search collects only developers who \tsuchida{state interest in GenAI} in their public profiles through one of the \tsuchida{two} English phrases. It misses \tsuchida{early adopters} who do not \tsuchida{state their interest} publicly, who describe themselves in other languages, or who use other vocabulary such as tool names. The results characterize this visible group of early adopters and should not be generalized to all developers who use \tsuchida{AI coding tools}. \tsuchida{Our selection criteria, a public statement of interest in GenAI, at least 10 public repositories, and at least 10 commits in the preceding year, may also correlate with enthusiasm for AI, open-source behavior, documentation activity, repository maturity, and project type, so the sample may differ from other early adopters in these respects. We also kept only repositories that the personal accounts of the early adopters own, and only English READMEs and issues, so the results may not hold for repositories owned by organizations or written in other languages.}

\tsuchida{The RQ2 results describe issue reporting in the 26 repositories that we analyze and do not generalize to the 531 early adopters. The 370 issues come from \tsuchida{222} issue authors, of whom the largest wrote 60 (16.2\%). We describe this concentration as a property of this sample, because our sample cannot distinguish whether it reflects how GenAI adoption is distributed among developers or results from our sampling. The RQ2 comparison rests on these 26 repositories, 13 in each group, so it can detect only large differences, and several confidence intervals are wide.}

\tsuchida{Removing the issues of the largest author does not change the RQ2 results. This author wrote all 60 issues in three repositories without ACFs. Removing them leaves 310 issues from 24 repositories, 13 with ACFs and 11 without ACFs. We fitted the same models to these issues. The model for Bug did not converge with the Laplace approximation, so we fitted all models with adaptive Gauss-Hermite quadrature with 10 points, which gives the same results as the Laplace approximation on the 370 issues. No value of Type or Outcome has an adjusted p value below 0.05 for ACFs (0.67 or higher). GenAI-related issues still concern External more often than Non-GenAI-related issues (odds ratio 4.31, adjusted p $<$ 0.001), and the share of GenAI-related issues shows no significant difference between the two repository groups (odds ratio 0.85, 95\% CI 0.18 to 4.07).}

\paragraph{Construct Validity}

We rely on two observable signals: profile keywords \tsuchida{that state interest in GenAI} (``AI Agent'' \tsuchida{and} ``Vibe Coding'') to identify \tsuchida{early adopters}, and \tsuchida{ACFs} (e.g., \texttt{.cursorrules}, \texttt{CLAUDE.md}) to identify \tsuchida{repositories with ACFs}. These configuration files originate from tools closer to the agent end of the assistant-to-agent spectrum. The same developers may also use AI Assistants through chat UIs, but chat use leaves no file in the repository, so our study captures only repository-visible AI use, which we interpret as an early step in the assistant-to-agent transition. \tsuchida{Conversely, an ACF shows only that the owner configured an AI coding tool for the repository; it does not show how much of the repository was developed with AI.}
\tsuchida{Repositories without ACFs may also be developed with AI tools that leave no file, such as chat assistants. Such AI use in repositories without ACFs may make the observed differences between the two repository groups smaller. We did not build the comparison group from repositories created before November 2022, because such a group could not separate the effect of AI-tool use from the effects of the period, such as tooling, README conventions, repository age, and project maturity, all of which affect the characteristics we measure.} \tsuchida{Finally, developing with AI tools and building software with GenAI features may be confounded. We labeled only the 26 repositories in RQ2 as GenAI-feature repositories or not. The 200 repositories in RQ1 are not labeled, so the RQ1 differences may partly reflect whether the software implements GenAI features rather than whether its owner configured an AI coding tool. In RQ2, the label is not significant for any value, but with 26 repositories the confidence intervals are wide (e.g., 0.63 to 7.54 for External), so our data cannot rule out such an effect.}

\section{Conclusion}
\label{sec:conclusion}

This section summarizes our findings and outlines future work on the transition from \tsuchida{Software Engineering 2.0 to Software Engineering 3.0~\cite{hassan2026se3}}.
\tsuchida{In the repositories of early adopters, our analysis shows modest distributional differences between repositories with and without ACFs in selected README structural characteristics, weak evidence of a difference in software categories, and no detected difference in their issues.} For RQ1.1, \tsuchida{the two repository groups differ in their category distributions, although no single category differs on its own}. For RQ1.2, \tsuchida{repositories with ACFs} have longer READMEs with more code blocks \tsuchida{and include a license more often}, while \tsuchida{repositories without ACFs} contain more URLs.

\tsuchida{For RQ2, we detect no difference in the Type or Outcome of issues between the two repository groups once GenAI relatedness is held fixed. External issues are more common among GenAI-related issues than among Non-GenAI-related issues, and we detect no association between External issues and ACFs. Because the comparison rests on 26 repositories, it can detect only large differences, and this result is not evidence that the two repository groups are the same.}
\tsuchida{Together, these findings describe the repositories and issues of early adopters in Software Engineering 2.0.}

Future research should compare this transition with fully agent-based development as autonomous agents mature. Other directions include examining how explicit signaling of GenAI usage affects repository popularity, analyzing AI usage patterns in multi-developer projects, and conducting longitudinal studies that track repository evolution.

\section*{Open Science and Data Availability}
Our replication package, containing the datasets, classification labels, and analysis scripts used in this study, is openly available at \url{https://doi.org/10.5281/zenodo.18302942}.

\section*{Declaration of Competing Interest}
Raula Gaikovina Kula is a member of the JSS Editorial Board. Given this role, he had no involvement in the peer review of this article and had no access to information regarding its peer review. The remaining authors declare that they have no known competing financial interests or personal relationships that could have appeared to influence the work reported in this paper.
\section*{Acknowledgments}
 This work was supported by JSPS KAKENHI Grant Number JP26K21197.

\section*{Declaration of generative AI and AI-assisted technologies in the manuscript preparation process}
During the preparation of this work, the authors used Claude to assist with drafting and refining the manuscript text, and to assist with writing code for data analysis. \tsuchida{One coder also used an LLM to draft initial labels for the issue classification and then reviewed and corrected every label manually (Section~\ref{sec:rq2}).} The authors reviewed and edited the output as needed and take full responsibility for the content of the published article.

\bibliographystyle{elsarticle-num}

\bibliography{references}

\end{document}

%% file: comments.tex
\usepackage{datenumber}
\usepackage{xspace}

\definecolor{chestnut}{rgb}{0.8, 0.36, 0.36}

\newcommand{\fixme}[1]{{\textcolor{blue}{#1}}\xspace}

\definecolor{chestnut}{rgb}{0.8, 0.36, 0.36}

\newcounter{dateone}\newcounter{datetwo}%
\newcommand{\daydifftoday}[3]{%
\setmydatenumber{dateone}{\the\year}{\the\month}{\the\day}%
\setmydatenumber{datetwo}{#1}{#2}{#3}%
\addtocounter{datetwo}{-\thedateone}%
\thedatetwo
}